\documentclass{article}
\usepackage[utf8]{inputenc}

\usepackage{listings}

\lstdefinestyle{matlabstyle}{
    language=Matlab,
    basicstyle=\ttfamily\footnotesize,
    keywordstyle=\color{blue},
    commentstyle=\color{green!50!black},
    stringstyle=\color{red},
    numbers=left,
    numberstyle=\tiny\color{gray},
    stepnumber=1,
    breaklines=true,
    captionpos=b,
    morekeywords={factorial, nchoosek, setdiff}
}

\usepackage{mathrsfs}
\usepackage{lmodern}
\usepackage[table,xcdraw]{xcolor}
\usepackage{colortbl}
\usepackage{graphicx}
\usepackage{multirow}%
\usepackage{amsmath,amssymb,amsfonts}%
\usepackage{amsthm}%
\usepackage[title]{appendix}%
\usepackage{float}
\usepackage{textcomp}%
\usepackage{manyfoot}%
\usepackage{booktabs}%
\usepackage{algorithm}%
\usepackage{algorithmicx}%
\usepackage{algpseudocode}%
\usepackage{listings}%
\usepackage{soul}
\usepackage{hyperref}
\usepackage{xurl}
\usepackage{caption}
\title{\Large \textbf {Optimizing PCA for Health and Care Research: A Reliable Approach to Component Selection}}

\makeatletter
\renewcommand\@date{{%
  \vspace{-\baselineskip}%
  \large\centering

  \begin{tabular}[t]{c@{\extracolsep{1em}}c} 
    \textbf{Nuwan Weeraratne} \\
      \normalsize Dept. of Mathematics and Statistics \\
      \normalsize University of Waikato, New Zealand \\
    \normalsize ncweera.nzinfo@gmail.com
  \end{tabular}%
    \begin{tabular}[t]{c@{\extracolsep{1em}}c} 
    \textbf{Lyn Hunt} \\
      \normalsize Dept. of Mathematics and Statistics \\
      \normalsize University of Waikato, New Zealand  \\
    \normalsize lah@waikato.ac.nz
  \end{tabular}

   \bigskip
  \begin{tabular}[t]{c@{\extracolsep{2em}}c} 
    \textbf{Jason Kurz} \\
      \normalsize Dept. of Mathematics and Statistics \\
      \normalsize University of Waikato, New Zealand \\
    \normalsize jason.kurz@waikato.ac.nz
  \end{tabular}

}}
\makeatother

\begin{document}

\maketitle

\begin{center}
\textbf{Abstract} \\
\end{center}

PCA is widely used in health and care research to analyze complex HD datasets, such as patient health records, genetic data, and medical imaging. By reducing dimensionality, PCA helps identify key patterns and trends, which can aid in disease diagnosis, treatment optimization, and the discovery of new biomarkers. However, the primary goal of any dimensional reduction technique is to reduce the dimensionality in a data set while keeping the essential information and variability. There are a few ways to do this in practice, such as the Kaiser-Guttman criterion, Cattell's Scree Test, and the percent cumulative variance approach. Unfortunately, the results of these methods are entirely different. That means using inappropriate methods to find the optimal number of PCs retained in PCA may lead to misinterpreted and inaccurate results in PCA and PCA-related health and care research applications. This contradiction becomes even more pronounced in HD settings where \textit{n} $<$ \textit{p}, making it even more critical to determine the best approach. Therefore, it is necessary to identify the issues of different techniques to select the optimal number of PCs retained in PCA.  Kaiser-Guttman criterion retains fewer PCs, causing overdispersion, while Cattell's scree test retains more PCs, compromising reliability. The percentage of cumulative variation criterion offers greater stability, consistently selecting the optimal number of components. Therefore, the Pareto chart, which shows both the cumulative percentage and the cut-off point for retained PCs, provides the most reliable method of selecting components, ensuring stability and enhancing PCA effectiveness, particularly in health-related research applications.\\


\section{Introduction}\label{sec1}

Over the last few years, due to the technology having evolved at a rapid pace, Big data analysis has become a trending topic in the Research and Development (R\&D) industry. In general, Big data leads to computational challenges, while analysis of high-dimensional data falls face to the curse of dimensionality. In healthcare research, large-scale patient data, genomic studies, and medical imaging generate vast amounts of high-dimensional data that require efficient processing. Without proper dimensionality reduction, these datasets become computationally expensive and challenging to analyze. To overcome this kind of issue, it is necessary to apply either regularization or dimensionality reduction techniques. Dimensionality Reduction refers to techniques that reduce the high dimensional space of original data into a low dimensional space for obtaining a set of principal variables. By reducing dimensionality, healthcare researchers can extract critical features from complex datasets, improving disease prediction models and enabling more accurate clinical decision-making. In other words, dimensionality reduction is a technique that we use to remove the least important information (sometimes redundant columns) from a data set. Moreover, dimensionality reduction can be divided into two parts namely feature selection and feature extraction. The prime difference between feature selection and feature extraction is that feature selection picks a subset from the original set of data and feature extraction generates a new set of features from the existing set of data. For instance, in medical imaging, reducing data dimensions enhances image processing algorithms, making diagnostics faster and more reliable. Similarly, in genomic studies, dimensionality reduction helps identify key genetic markers associated with diseases, aiding in precision medicine. If all the features are equally relevant for the study, the Principal Component Analysis (PCA) is the most suitable statistical technique to reduce dimensionality and eliminate redundancy. \\

PCA is used to reduce the dimensionality of a data set consisting of many variables correlated with each other, either heavily or lightly, while retaining the variation present in the dataset to the maximum extent. This process involves transforming the original variables into a new set of variables, called principal components (PCs), which are orthogonal and arranged in descending order of variance, ensuring that each successive component captures less variation from the original data. In healthcare research, PCA plays a crucial role in analyzing high-throughput biological data, such as gene expression profiles, where thousands of genes are measured simultaneously. By reducing the number of dimensions, PCA helps researchers identify significant genetic patterns that may be linked to diseases. Thus, the first principal component captures the maximum variation from the original variables. Similarly, in patient health records, where multiple clinical variables are recorded for each individual, PCA aids in summarizing essential health indicators, leading to more efficient patient stratification and personalized treatment approaches. Since principal components are the eigenvectors of the covariance matrix, they are inherently orthogonal. \\

By simplifying these high-dimensional datasets, PCA enhances predictive modeling and medical decision-making. However, the primary goal of any dimensionality reduction technique is to reduce the dimensionality in a dataset while keeping the essential information and variability. There are a few ways to do this in practice, such as the Kaiser-Guttman Criterion, Cattell's Scree Test, and the percent of cumulative variance approach.\\

\pmb{Kaiser-Guttman Criterion:} This approach recommends excluding components with eigenvalues of 1.00 or less, as they explain less variation than the original standardized variables. Highlighting components with higher eigenvalues emphasizes the most significant parts of the data, thereby improving the clarity and effectiveness of PCA. It determines the number of components to retain based on the equation involving the eigenvalue $\lambda$. However, this method can be problematic: it tends to select too many components when there are many variables and too few when there are few variables. This variability makes it less reliable across different dimensional settings \cite{Guttman_1954}.\\

\pmb{Cattell's Scree Test:} This involves plotting the eigenvalues against the component numbers and identifying where the plot levels off, resembling a scree slope. The test requires visual inspection to determine the cutoff point where the plot starts to flatten. While it works well with solid factors, it suffers from subjectivity and lacks a clear definition of the cutoff point. This ambiguity is particularly problematic when there are no obvious breaks or multiple apparent breaks in the plot, making it challenging to apply consistently in both \textit{n} $<$ \textit{p} and \textit{n} $>$ \textit{p} high dimensional scenarios \cite{Cattell01041966}.\\

\pmb{Percent of Cumulative Variance:} This approach retains enough components to explain a specific percentage of the total variance, typically 70-80$\%$. The method is straightforward: it involves summing all eigenvalues and retaining the components that cumulatively account for the desired percentage of variance. However, it has been criticized for its subjectivity due to the arbitrary nature of the chosen variance threshold. This can lead to retaining too many or too few components, especially in high-dimensional settings \textit{n} $<$ \textit{p}, where the proportion of variance explained by individual components can be more dispersed.\\

However, selecting the optimal number of PCs to retain is critical, as it directly affects the quality of the results. Retaining too few PCs may lead to the loss of important medical information, while keeping too many may introduce noise and redundancy. For example, selecting the right number of PCs in disease classification models ensures that only the most relevant clinical characteristics are retained, improving accuracy and interpretability. By carefully choosing the retained PCs, researchers can balance dimensionality reduction with information preservation, ultimately leading to better insights and more reliable healthcare applications.

\section{Problem Statement}\label{sec2}

Conspicuously, there are a number of different ways to approach the issue of deciding how many components or factors to include in PCA. However, the main drawback is that each of these methods (Kaiser-Guttman Criterion, Cattell’s Scree Test, and Percent of Cumulative Variance) often provides contradictory solutions when compared to others. This raises the question: which method is truly correct?

\begin{figure}[h]
\centering
\includegraphics[width=1.0\textwidth]{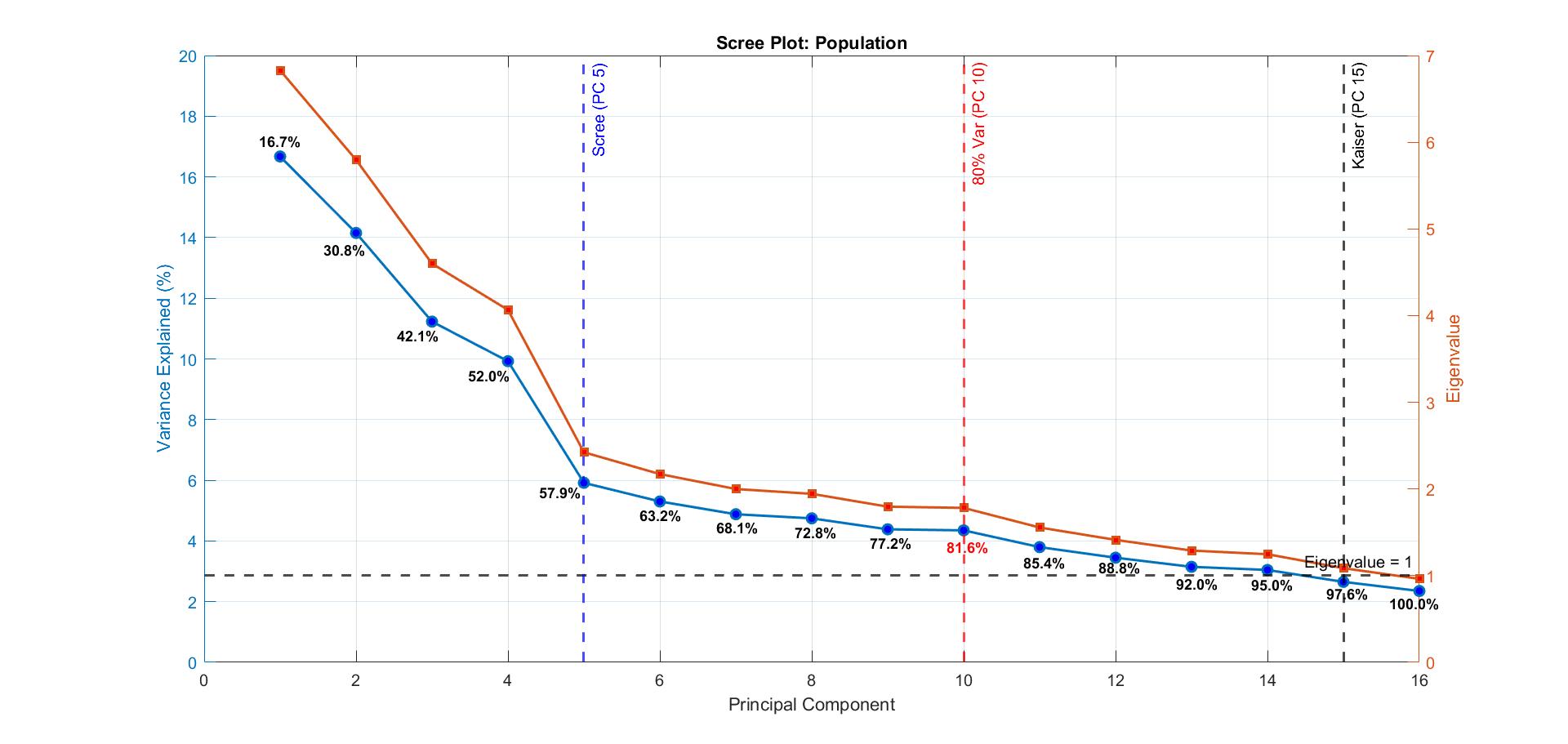}
    \caption{Disagreement in PC Selection Methods}
    \caption*{Note: The data source is \href{https://www.spss-tutorials.com/downloads/dole-survey.sav} {survey.sav} \cite{spsstutorialsSPSSFactor}, based on a survey of 388 applicants.}
\label{fig1}
\end{figure}

As shown in Figure 1, the Kaiser-Guttman Criterion selects too many PCs, while Cattell’s Scree Test selects too few. The Percent of Cumulative Variance approach falls in between, selecting a moderate number. This highlights the contradiction among these methods in determining the number of PCs to retain. Such contradictions can lead to misinterpretations and inaccuracies in PCA, particularly in healthcare research applications. If this occurs in healthcare research, misleading statistics can lead to critical errors, potentially affecting diagnoses, treatments, and policy decisions. Therefore, addressing this issue is crucial to prevent misleading statistical conclusions and ensure reliable analysis.

\section{Methodology}

This study investigates contradictions among different methods for determining the number of principal components (PCs) to retain through a structured simulation analysis. The data are generated from a 10-dimensional ($p = 10$) multivariate normal distribution, where the mean vector is defined as $\boldsymbol{\mu} = \mathbf{0}_{10}$, and the covariance matrix $\boldsymbol{\Sigma}$ is constructed as a positive semi-definite matrix, ensuring valid dependence structures among variables.\\

For each simulation run, a sample of size $n$ is drawn from this distribution, considering a range of sample sizes $n \in \{2, 3, 4, 5, 10, 15, 20, 30, 40, 50, 100\}$. These values correspond to different dimensionality ratios ($p/n$), ranging from extreme high-dimensional settings ($p/n = 5.00$ when $n=2$) to more balanced cases ($p/n = 0.10$ when $n=100$). To account for sampling variability, the process is repeated over 100 independent iterations for each sample size. Principal component analysis (PCA) is then applied to each sample, and the number of retained PCs is determined using three commonly used criteria: the Kaiser-Guttman Criterion, which retains components with eigenvalues greater than 1; Cattell’s Scree Test, which identifies the "elbow" where eigenvalues exhibit the largest rate of decline; and the Percent of Cumulative Variance approach, which selects the minimal number of PCs explaining at least 80\% of the total variance.\\

To assess statistical differences among these selection methods, a one-way analysis of variance is performed at a 5\% significance level ($\alpha = 0.05$) with a 95\% confidence level. This test evaluates whether the number of retained PCs differs significantly across methods for various sample sizes, thereby quantifying the contradictions observed in PC selection.\\

Given that PCA estimation becomes problematic in high-dimensional settings where $n < p$, alternative covariance estimation techniques are incorporated to improve stability. Instead of relying solely on maximum likelihood estimation (MLE), which may yield unreliable estimates under small-sample conditions, two alternative methods are applied: the Ledoit-Wolf Estimator \cite{ledoit2004well} and the Pairwise Differences Covariance Estimation \cite{weeraratne2025principalcomponentanalysisn}.\\

The study aims to highlight the inconsistencies among these PCA selection methods and the impact of sample size on their reliability. Addressing these contradictions is particularly critical in fields such as healthcare research, where incorrect statistical interpretations can lead to misleading conclusions with serious implications. By integrating alternative covariance estimators and statistical validation techniques, this study provides a structured framework to enhance PCA-based decision-making in both standard and high-dimensional contexts.

\section {Results and Discussion}

As shown in Table 1, the analysis examines the number of PCs retained by three selection criteria: Kaiser-Guttman Criterion, Cattell’s Scree Test, and the Percent of Cumulative Variance (80\%), across population and sample levels under varying sample sizes \( n \) with a fixed dimensionality of \( p = 10 \). In the population, the Kaiser-Guttman Criterion consistently selects 8 PCs, Cattell’s Scree Test selects 1 PC, and the Percent of Cumulative Variance method selects 4 PCs. However, in the sample, the retained PCs vary depending on sample size.\\

Table 1: Number of Principal Components Retained Under Different Criteria
\begin{table}[ht]
\resizebox{\columnwidth}{!}{%
\begin{tabular}{@{}|c|c|c|cccccc|@{}}
\toprule
\rowcolor[HTML]{C0C0C0} 
\cellcolor[HTML]{C0C0C0} & \cellcolor[HTML]{C0C0C0} & \cellcolor[HTML]{C0C0C0} & \multicolumn{6}{c|}{\cellcolor[HTML]{C0C0C0}No: of Principal Components to retain} \\ \cmidrule(l){4-9} 
\rowcolor[HTML]{C0C0C0} 
\cellcolor[HTML]{C0C0C0} & \cellcolor[HTML]{C0C0C0} & \cellcolor[HTML]{C0C0C0} & \multicolumn{3}{c|}{\cellcolor[HTML]{C0C0C0}Population} & \multicolumn{3}{c|}{\cellcolor[HTML]{C0C0C0}Sample} \\ \cmidrule(l){4-9} 
\rowcolor[HTML]{C0C0C0} 
\multirow{-3}{*}{\cellcolor[HTML]{C0C0C0}p} & \multirow{-3}{*}{\cellcolor[HTML]{C0C0C0}n} & \multirow{-3}{*}{\cellcolor[HTML]{C0C0C0}p/n} & \multicolumn{1}{c|}{\cellcolor[HTML]{C0C0C0}\begin{tabular}[c]{@{}c@{}}Kaiser\\ Guttman \\ Criterion\end{tabular}} & \multicolumn{1}{c|}{\cellcolor[HTML]{C0C0C0}\begin{tabular}[c]{@{}c@{}}Cattell’s \\ Scree \\ Test\end{tabular}} & \multicolumn{1}{c|}{\cellcolor[HTML]{C0C0C0}\begin{tabular}[c]{@{}c@{}}Percent of\\ Cumulative\\ Variance (80\%)\end{tabular}} & \multicolumn{1}{c|}{\cellcolor[HTML]{C0C0C0}\begin{tabular}[c]{@{}c@{}}Kaiser\\ Guttman\\ Criterion\end{tabular}} & \multicolumn{1}{c|}{\cellcolor[HTML]{C0C0C0}\begin{tabular}[c]{@{}c@{}}Cattell's\\ Scree\\ Test\end{tabular}} & \begin{tabular}[c]{@{}c@{}}Percent of\\ Cumulative\\ Variance (80\%)\end{tabular} \\ \midrule
10 & 100 & 0.10 & \multicolumn{1}{c|}{8} & \multicolumn{1}{c|}{1} & \multicolumn{1}{c|}{4} & \multicolumn{1}{c|}{8} & \multicolumn{1}{c|}{1} & 4 \\ \midrule
10 & 50 & 0.20 & \multicolumn{1}{c|}{8} & \multicolumn{1}{c|}{1} & \multicolumn{1}{c|}{4} & \multicolumn{1}{c|}{8} & \multicolumn{1}{c|}{1} & 4 \\ \midrule
10 & 40 & 0.25 & \multicolumn{1}{c|}{8} & \multicolumn{1}{c|}{1} & \multicolumn{1}{c|}{4} & \multicolumn{1}{c|}{8} & \multicolumn{1}{c|}{1} & 4 \\ \midrule
10 & 30 & 0.33 & \multicolumn{1}{c|}{8} & \multicolumn{1}{c|}{1} & \multicolumn{1}{c|}{4} & \multicolumn{1}{c|}{8} & \multicolumn{1}{c|}{1} & 4 \\ \midrule
10 & 20 & 0.50 & \multicolumn{1}{c|}{8} & \multicolumn{1}{c|}{1} & \multicolumn{1}{c|}{4} & \multicolumn{1}{c|}{8} & \multicolumn{1}{c|}{1} & 3 \\ \midrule
10 & 15 & 0.67 & \multicolumn{1}{c|}{8} & \multicolumn{1}{c|}{1} & \multicolumn{1}{c|}{4} & \multicolumn{1}{c|}{7} & \multicolumn{1}{c|}{1} & 3 \\ \midrule
10 & 10 & 1.00 & \multicolumn{1}{c|}{8} & \multicolumn{1}{c|}{1} & \multicolumn{1}{c|}{4} & \multicolumn{1}{c|}{6} & \multicolumn{1}{c|}{1} & 3 \\ \midrule
10 & 5 & 2.00 & \multicolumn{1}{c|}{8} & \multicolumn{1}{c|}{1} & \multicolumn{1}{c|}{4} & \multicolumn{1}{c|}{4} & \multicolumn{1}{c|}{1} & 2 \\ \midrule
10 & 4 & 2.50 & \multicolumn{1}{c|}{8} & \multicolumn{1}{c|}{1} & \multicolumn{1}{c|}{4} & \multicolumn{1}{c|}{3} & \multicolumn{1}{c|}{1} & 2 \\ \midrule
10 & 3 & 3.33 & \multicolumn{1}{c|}{8} & \multicolumn{1}{c|}{1} & \multicolumn{1}{c|}{4} & \multicolumn{1}{c|}{2} & \multicolumn{1}{c|}{1} & 2 \\ \midrule
10 & 2 & 5.00 & \multicolumn{1}{c|}{8} & \multicolumn{1}{c|}{1} & \multicolumn{1}{c|}{4} & \multicolumn{1}{c|}{1} & \multicolumn{1}{c|}{1} & 1 \\ \bottomrule
\end{tabular}%
}
\end{table}

For larger samples (\( n \geq 30 \)), the sample-based PC selection remains nearly identical to the population results. However, as \( n \) decreases, deviations become more evident, particularly for the Kaiser-Guttman Criterion and the Percent of Cumulative Variance approach. When the sample size is moderate (\( n = 10 \) to \( n = 30 \)), the Kaiser-Guttman Criterion starts selecting fewer PCs in the sample than in the population, dropping from 8 to 6 when \( n = 10 \). As the sample size further decreases (\( n < 10 \)), instability becomes more prominent, with the Kaiser-Guttman and Percent of Cumulative Variance methods showing sharp reductions in retained PCs.\\

At extreme cases, when \( p/n \geq 2.0 \) (i.e., \( n \leq 5 \)), the retained number of PCs is significantly lower than in the population, indicating the impact of high-dimensionality effects. At \( p/n = 5.0 \) (\( n = 2 \)), both the Kaiser-Guttman and Percent of Cumulative Variance methods reduce their selection to only one PC, revealing substantial sensitivity to small samples. Cattell’s Scree Test, in contrast, remains stable, selecting only 1 PC in all cases, making it overly conservative and potentially leading to oversimplification.\\

\begin{table}[h]
\centering
\renewcommand{\thetable}{2} 
\begin{tabular}{lccccc}
\toprule
Source & SS & df & MS & F & Prob $>$ F \\ 
\midrule
Groups & 124.424 & 2  & 62.2121 & 21.93 & 1.34901e-06 \\ 
Error  & 85.091  & 30 & 2.8364  &      &             \\ 
Total  & 209.515 & 32 &         &      &             \\ 
\bottomrule
\end{tabular}
\caption{ANOVA Test Results}
\label{tab:anova_results}
\end{table}

The one-way ANOVA test in Table 2 evaluates differences in the number of PCs retained between selection methods. The highly significant \textit{F}-statistic (\textit{F} = 21.93, \textit{p} $<$ 0.000001) suggests that at least one method differs significantly from the others.\\

Table 3: Pairwise multiple comparisons
\begin{table}[ht]
\resizebox{\columnwidth}{!}{%
\begin{tabular}{@{}cccccc@{}}
\toprule
Comparison & \begin{tabular}[c]{@{}c@{}}Mean\\ Difference\\ (a)\end{tabular} & \begin{tabular}[c]{@{}c@{}}95\% CI\\ Lower\\ Band\end{tabular} & \begin{tabular}[c]{@{}c@{}}95\% CI\\ Upper\\ Band\end{tabular} & p-value & \begin{tabular}[c]{@{}c@{}}Significance\\ (p\textless{}0.05)\end{tabular} \\ \midrule
KGC vs. CSP & 2.9569 & 4.7273 & 6.4976 & 0.0000 & yes \\ \midrule
KGC vs. CV & 1.0478 & 2.8182 & 4.5886 & 0.0013 & yes \\ \midrule
CSP vs. CV & -3.6795 & -1.9091 & -0.1387 & 0.0325 & yes \\ \bottomrule
\end{tabular}%
}
\end{table}

According to Table 3, the results of Tukey's HSD multiple comparisons test revealed that there are statistically significant differences between all three PCA retention methods: Kaiser-Guttman Criterion (KGC), Scree Plot (CSP), and Cumulative Variance (CV). The comparison between the Kaiser-Guttman and Scree-Plot methods shows a significant mean difference of 2.9569, with a p-value of 0.0000, indicating that the two methods yield different results. Similarly, the Kaiser-Guttman method also significantly differs from the Cumulative Variance method, with a mean difference of 1.0478 and a p-value of 0.0013. Lastly, the Scree-Plot method significantly differs from the Cumulative Variance method, with a mean difference of -3.6795 and a p-value of 0.0325. These results suggest that the methods do not perform equivalently, and each offers distinct results in terms of PCA retention, making the differences meaningful and not due to random chance.\\

As previously discussed, when the ratio of \( p/n \geq 1 \) (indicating a high-dimensional scenario where \( n < p \)), the number of retained PCs is significantly lower than in the population, highlighting the impact of high-dimensionality effects. This discrepancy arises because the maximum likelihood sample covariance estimation becomes ill-conditioned in \( n < p \) high-dimensional settings. However, as Nuwan et al. \cite{weeraratne2025principalcomponentanalysisn} demonstrated with their novel Pairwise Differences Covariance Estimation, this approach offers a reasonable solution for such \( n < p \) applications.\\

In high-dimensional settings where \( n < p \), the novel regularized PDC estimators, particularly the Standardized PDC methods, demonstrate superior performance compared to traditional approaches \cite{weeraratne2025principalcomponentanalysisn}. They achieve lower cosine similarity error for the first principal component and exhibit reduced overdispersion error. Notably, Standardized PDC outperforms both the conventional maximum likelihood covariance estimation and its leading alternative, the Ledoit-Wolf covariance estimator. To illustrate how these different methods influence the selection of the number of principal components, we employ the percentage of cumulative variance criterion, as it consistently provides the most realistic solutions across various scenarios.\\

Table 4: Cumulative Variance Explained by PCs for Different Estimation Methods
\begin{table}[ht]
\resizebox{\columnwidth}{!}{%
\begin{tabular}{@{}|c|c|ccc|ccc|ccc|ccc|@{}}
\toprule
\rowcolor[HTML]{C0C0C0} 
\cellcolor[HTML]{C0C0C0} & \cellcolor[HTML]{C0C0C0} & \multicolumn{3}{c|}{\cellcolor[HTML]{C0C0C0}\textbf{Population}} & \multicolumn{3}{c|}{\cellcolor[HTML]{C0C0C0}\textbf{Sample (MLE)}} & \multicolumn{3}{c|}{\cellcolor[HTML]{C0C0C0}\textbf{Sample (LW)}} & \multicolumn{3}{c|}{\cellcolor[HTML]{C0C0C0}\textbf{Sample (SPDC)}} \\ \cmidrule(l){3-14} 
\rowcolor[HTML]{C0C0C0} 
\multirow{-2}{*}{\cellcolor[HTML]{C0C0C0}\textbf{p}} & \multirow{-2}{*}{\cellcolor[HTML]{C0C0C0}\textbf{n}} & \multicolumn{1}{c|}{\cellcolor[HTML]{C0C0C0}\textbf{ID}} & \multicolumn{1}{c|}{\cellcolor[HTML]{C0C0C0}\textbf{\begin{tabular}[c]{@{}c@{}}Cum. \\ Var\end{tabular}}} & \textbf{\begin{tabular}[c]{@{}c@{}}\# of \\ PCs\end{tabular}} & \multicolumn{1}{c|}{\cellcolor[HTML]{C0C0C0}\textbf{ID}} & \multicolumn{1}{c|}{\cellcolor[HTML]{C0C0C0}\textbf{\begin{tabular}[c]{@{}c@{}}Cum. \\ Var\end{tabular}}} & \textbf{\begin{tabular}[c]{@{}c@{}}\# of \\ PCs\end{tabular}} & \multicolumn{1}{c|}{\cellcolor[HTML]{C0C0C0}\textbf{ID}} & \multicolumn{1}{c|}{\cellcolor[HTML]{C0C0C0}\textbf{\begin{tabular}[c]{@{}c@{}}Cum. \\ Var\end{tabular}}} & \textbf{\begin{tabular}[c]{@{}c@{}}\# of \\ PCs\end{tabular}} & \multicolumn{1}{c|}{\cellcolor[HTML]{C0C0C0}\textbf{ID}} & \multicolumn{1}{c|}{\cellcolor[HTML]{C0C0C0}\textbf{\begin{tabular}[c]{@{}c@{}}Cum. \\ Var\end{tabular}}} & \textbf{\begin{tabular}[c]{@{}c@{}}\# of \\ PCs\end{tabular}} \\ \midrule
 &  & \multicolumn{1}{c|}{PC-1} & \multicolumn{1}{c|}{39.06} &  & \multicolumn{1}{c|}{PC-1} & \multicolumn{1}{c|}{60.19} &  & \multicolumn{1}{c|}{PC-1} & \multicolumn{1}{c|}{37.19} &  & \multicolumn{1}{c|}{PC-1} & \multicolumn{1}{c|}{50.09} &  \\ \cmidrule(lr){3-4} \cmidrule(lr){6-7} \cmidrule(lr){9-10} \cmidrule(lr){12-13}
 &  & \multicolumn{1}{c|}{PC-2} & \multicolumn{1}{c|}{60.75} &  & \multicolumn{1}{c|}{PC-2} & \multicolumn{1}{c|}{85.98} &  & \multicolumn{1}{c|}{PC-2} & \multicolumn{1}{c|}{55.05} &  & \multicolumn{1}{c|}{PC-2} & \multicolumn{1}{c|}{79.16} &  \\ \cmidrule(lr){3-4} \cmidrule(lr){6-7} \cmidrule(lr){9-10} \cmidrule(lr){12-13}
 &  & \multicolumn{1}{c|}{PC-3} & \multicolumn{1}{c|}{75.81} &  & \multicolumn{1}{c|}{PC-3} & \multicolumn{1}{c|}{96.29} &  & \multicolumn{1}{c|}{PC-3} & \multicolumn{1}{c|}{64.89} &  & \multicolumn{1}{c|}{PC-3} & \multicolumn{1}{c|}{94.19} &  \\ \cmidrule(lr){3-4} \cmidrule(lr){6-7} \cmidrule(lr){9-10} \cmidrule(lr){12-13}
\multirow{-4}{*}{10} & \multirow{-4}{*}{5} & \multicolumn{1}{c|}{PC-4} & \multicolumn{1}{c|}{83.58} & \multirow{-4}{*}{4} & \multicolumn{1}{c|}{PC-4} & \multicolumn{1}{c|}{100} & \multirow{-4}{*}{2} & \multicolumn{1}{c|}{PC-4} & \multicolumn{1}{c|}{71.48} & \multirow{-4}{*}{6} & \multicolumn{1}{c|}{PC-4} & \multicolumn{1}{c|}{100} & \multirow{-4}{*}{3} \\ \midrule
 &  & \multicolumn{1}{c|}{PC-1} & \multicolumn{1}{c|}{39.06} &  & \multicolumn{1}{c|}{PC-1} & \multicolumn{1}{c|}{57.25} &  & \multicolumn{1}{c|}{PC-1} & \multicolumn{1}{c|}{34.98} &  & \multicolumn{1}{c|}{PC-1} & \multicolumn{1}{c|}{46.30} &  \\ \cmidrule(lr){3-4} \cmidrule(lr){6-7} \cmidrule(lr){9-10} \cmidrule(lr){12-13}
 &  & \multicolumn{1}{c|}{PC-2} & \multicolumn{1}{c|}{60.75} &  & \multicolumn{1}{c|}{PC-2} & \multicolumn{1}{c|}{82.04} &  & \multicolumn{1}{c|}{PC-2} & \multicolumn{1}{c|}{52.23} &  & \multicolumn{1}{c|}{PC-2} & \multicolumn{1}{c|}{73.96} &  \\ \cmidrule(lr){3-4} \cmidrule(lr){6-7} \cmidrule(lr){9-10} \cmidrule(lr){12-13}
 &  & \multicolumn{1}{c|}{PC-3} & \multicolumn{1}{c|}{75.81} &  & \multicolumn{1}{c|}{PC-3} & \multicolumn{1}{c|}{93.99} &  & \multicolumn{1}{c|}{PC-3} & \multicolumn{1}{c|}{62.93} &  & \multicolumn{1}{c|}{PC-3} & \multicolumn{1}{c|}{89.75} &  \\ \cmidrule(lr){3-4} \cmidrule(lr){6-7} \cmidrule(lr){9-10} \cmidrule(lr){12-13}
\multirow{-4}{*}{10} & \multirow{-4}{*}{6} & \multicolumn{1}{c|}{PC-4} & \multicolumn{1}{c|}{83.58} & \multirow{-4}{*}{4} & \multicolumn{1}{c|}{PC-4} & \multicolumn{1}{c|}{98.43} & \multirow{-4}{*}{2} & \multicolumn{1}{c|}{PC-4} & \multicolumn{1}{c|}{70.00} & \multirow{-4}{*}{6} & \multicolumn{1}{c|}{PC-4} & \multicolumn{1}{c|}{97.28} & \multirow{-4}{*}{3} \\ \midrule
 &  & \multicolumn{1}{c|}{PC-1} & \multicolumn{1}{c|}{39.06} &  & \multicolumn{1}{c|}{PC-1} & \multicolumn{1}{c|}{56.13} &  & \multicolumn{1}{c|}{PC-1} & \multicolumn{1}{c|}{34.62} &  & \multicolumn{1}{c|}{PC-1} & \multicolumn{1}{c|}{45.18} &  \\ \cmidrule(lr){3-4} \cmidrule(lr){6-7} \cmidrule(lr){9-10} \cmidrule(lr){12-13}
 &  & \multicolumn{1}{c|}{PC-2} & \multicolumn{1}{c|}{60.75} &  & \multicolumn{1}{c|}{PC-2} & \multicolumn{1}{c|}{80.41} &  & \multicolumn{1}{c|}{PC-2} & \multicolumn{1}{c|}{51.53} &  & \multicolumn{1}{c|}{PC-2} & \multicolumn{1}{c|}{71.08} &  \\ \cmidrule(lr){3-4} \cmidrule(lr){6-7} \cmidrule(lr){9-10} \cmidrule(lr){12-13}
 &  & \multicolumn{1}{c|}{PC-3} & \multicolumn{1}{c|}{75.81} &  & \multicolumn{1}{c|}{PC-3} & \multicolumn{1}{c|}{91.74} &  & \multicolumn{1}{c|}{PC-3} & \multicolumn{1}{c|}{61.90} &  & \multicolumn{1}{c|}{PC-3} & \multicolumn{1}{c|}{86.65} &  \\ \cmidrule(lr){3-4} \cmidrule(lr){6-7} \cmidrule(lr){9-10} \cmidrule(lr){12-13}
\multirow{-4}{*}{10} & \multirow{-4}{*}{7} & \multicolumn{1}{c|}{PC-4} & \multicolumn{1}{c|}{83.58} & \multirow{-4}{*}{4} & \multicolumn{1}{c|}{PC-4} & \multicolumn{1}{c|}{96.97} & \multirow{-4}{*}{2} & \multicolumn{1}{c|}{PC-4} & \multicolumn{1}{c|}{69.32} & \multirow{-4}{*}{6} & \multicolumn{1}{c|}{PC-4} & \multicolumn{1}{c|}{95.04} & \multirow{-4}{*}{3} \\ \bottomrule
\end{tabular}%
}
\end{table}

As shown in Table 4, under high-dimensional settings where \( n < p \), traditional maximum likelihood estimation tends to overestimate the variance of the first principal component, leading to the selection of fewer components than expected from the true population distribution. This bias stems from the ill-conditioned nature of the covariance estimator, which inflates the dominance of the first principal component. In contrast, the Ledoit-Wolf covariance estimator attempts to mitigate this overdispersion but instead underestimates the first principal component while also suppressing the variance of subsequent components, resulting in an excessive number of retained principal components, which undermines the goal of dimensionality reduction. The newly proposed Standardized PDC method effectively addresses these limitations by minimizing estimation errors and providing more accurate variance estimates, ensuring a balanced selection of principal components that better align with the underlying data structure.\\

\textbf{Experiments on Public Data:} This data set is derived from a study that examined gene expression in two types of acute leukemia: acute lymphoblastic leukemia (ALL) and acute myeloid leukemia (AML). Gene expression levels were assessed using Affymetrix high-density oligonucleotide arrays, which cover 6817 human genes. The dataset includes 25 observations across three leukemia subtypes: 13 cases of B-cell ALL, 7 cases of AML, and 5 cases of T-cell ALL. Gene expression data is provided for 40 genes, labeled Gene1 to Gene40 \cite{dudoit2002comparison}. 

\newpage

Table 5 illustrates the number of PCs to retain according to different methods. In our experiments with synthetic data, both the  Kaiser-Guttman Criterion and Cattell’s Scree Test methods suggested impractical numbers of PCs. For instance, Cattell’s Scree Test method recommended retaining only 2 PCs, which explained just 37\% of the total variance, undermining the primary goal of dimensionality reduction. In contrast, the percentage of cumulative variance method provided a more realistic and reasonable approach to dimensionality reduction for the given percentage.\\

Table 5: No: of PCs to be retained | Leukemia Data
\begin{table}[ht]
\resizebox{\columnwidth}{!}{%
\begin{tabular}{@{}|c|c|c|c|ccc|ccccccccc|@{}}
\toprule
\rowcolor[HTML]{C0C0C0} 
\cellcolor[HTML]{C0C0C0} & \cellcolor[HTML]{C0C0C0} & \cellcolor[HTML]{C0C0C0} & \cellcolor[HTML]{C0C0C0} & \multicolumn{3}{c|}{\cellcolor[HTML]{C0C0C0}} & \multicolumn{9}{c|}{\cellcolor[HTML]{C0C0C0}\textbf{\begin{tabular}[c]{@{}c@{}}\# of Sample PCs\\ (Based on 80\% CV method)\end{tabular}}} \\ \cmidrule(l){8-16} 
\rowcolor[HTML]{C0C0C0} 
\cellcolor[HTML]{C0C0C0} & \cellcolor[HTML]{C0C0C0} & \cellcolor[HTML]{C0C0C0} & \cellcolor[HTML]{C0C0C0} & \multicolumn{3}{c|}{\multirow{-2}{*}{\cellcolor[HTML]{C0C0C0}\textbf{\begin{tabular}[c]{@{}c@{}}\# of \\ Population PCs\end{tabular}}}} & \multicolumn{3}{c|}{\cellcolor[HTML]{C0C0C0}\textbf{MLE}} & \multicolumn{3}{c|}{\cellcolor[HTML]{C0C0C0}\textbf{LW}} & \multicolumn{3}{c|}{\cellcolor[HTML]{C0C0C0}\textbf{SPDC}} \\ \cmidrule(l){5-16} 
\rowcolor[HTML]{C0C0C0} 
\multirow{-3}{*}{\cellcolor[HTML]{C0C0C0}\textbf{p}} & \multirow{-3}{*}{\cellcolor[HTML]{C0C0C0}\textbf{n}} & \multirow{-3}{*}{\cellcolor[HTML]{C0C0C0}\textbf{ID}} & \multirow{-3}{*}{\cellcolor[HTML]{C0C0C0}\textbf{CV}} & \multicolumn{1}{c|}{\cellcolor[HTML]{C0C0C0}\textbf{KGC}} & \multicolumn{1}{c|}{\cellcolor[HTML]{C0C0C0}\textbf{CST}} & \textbf{\begin{tabular}[c]{@{}c@{}}80\%\\ CV\end{tabular}} & \multicolumn{1}{c|}{\cellcolor[HTML]{C0C0C0}\textbf{ID}} & \multicolumn{1}{c|}{\cellcolor[HTML]{C0C0C0}\textbf{CV}} & \multicolumn{1}{c|}{\cellcolor[HTML]{C0C0C0}\textbf{\begin{tabular}[c]{@{}c@{}}\# of\\ PCs\end{tabular}}} & \multicolumn{1}{c|}{\cellcolor[HTML]{C0C0C0}\textbf{ID}} & \multicolumn{1}{c|}{\cellcolor[HTML]{C0C0C0}\textbf{CV}} & \multicolumn{1}{c|}{\cellcolor[HTML]{C0C0C0}\textbf{\begin{tabular}[c]{@{}c@{}}\# of\\ PCs\end{tabular}}} & \multicolumn{1}{c|}{\cellcolor[HTML]{C0C0C0}\textbf{ID}} & \multicolumn{1}{c|}{\cellcolor[HTML]{C0C0C0}\textbf{CV}} & \textbf{\begin{tabular}[c]{@{}c@{}}\# of\\ PCs\end{tabular}} \\ \midrule
 &  & PC-1 & 21.11 & \multicolumn{1}{c|}{} & \multicolumn{1}{c|}{} &  & \multicolumn{1}{c|}{PC-1} & \multicolumn{1}{c|}{27.52} & \multicolumn{1}{c|}{} & \multicolumn{1}{c|}{PC-1} & \multicolumn{1}{c|}{18.81} & \multicolumn{1}{c|}{} & \multicolumn{1}{c|}{PC-1} & \multicolumn{1}{c|}{22.94} &  \\ \cmidrule(lr){3-4} \cmidrule(lr){8-9} \cmidrule(lr){11-12} \cmidrule(lr){14-15}
 &  & PC-2 & 37.34 & \multicolumn{1}{c|}{} & \multicolumn{1}{c|}{} &  & \multicolumn{1}{c|}{PC-2} & \multicolumn{1}{c|}{45.09} & \multicolumn{1}{c|}{} & \multicolumn{1}{c|}{PC-2} & \multicolumn{1}{c|}{30.93} & \multicolumn{1}{c|}{} & \multicolumn{1}{c|}{PC-2} & \multicolumn{1}{c|}{38.96} &  \\ \cmidrule(lr){3-4} \cmidrule(lr){8-9} \cmidrule(lr){11-12} \cmidrule(lr){14-15}
 &  & PC-3 & 47.58 & \multicolumn{1}{c|}{} & \multicolumn{1}{c|}{} &  & \multicolumn{1}{c|}{PC-3} & \multicolumn{1}{c|}{57.36} & \multicolumn{1}{c|}{} & \multicolumn{1}{c|}{PC-3} & \multicolumn{1}{c|}{39.63} & \multicolumn{1}{c|}{} & \multicolumn{1}{c|}{PC-3} & \multicolumn{1}{c|}{50.05} &  \\ \cmidrule(lr){3-4} \cmidrule(lr){8-9} \cmidrule(lr){11-12} \cmidrule(lr){14-15}
 &  & PC-4 & 54.82 & \multicolumn{1}{c|}{} & \multicolumn{1}{c|}{} &  & \multicolumn{1}{c|}{PC-4} & \multicolumn{1}{c|}{66.18} & \multicolumn{1}{c|}{} & \multicolumn{1}{c|}{PC-4} & \multicolumn{1}{c|}{46.14} & \multicolumn{1}{c|}{} & \multicolumn{1}{c|}{PC-4} & \multicolumn{1}{c|}{58.68} &  \\ \cmidrule(lr){3-4} \cmidrule(lr){8-9} \cmidrule(lr){11-12} \cmidrule(lr){14-15}
 &  & PC-5 & 61.60 & \multicolumn{1}{c|}{} & \multicolumn{1}{c|}{} &  & \multicolumn{1}{c|}{PC-5} & \multicolumn{1}{c|}{72.69} & \multicolumn{1}{c|}{} & \multicolumn{1}{c|}{PC-5} & \multicolumn{1}{c|}{51.16} & \multicolumn{1}{c|}{} & \multicolumn{1}{c|}{PC-5} & \multicolumn{1}{c|}{65.90} &  \\ \cmidrule(lr){3-4} \cmidrule(lr){8-9} \cmidrule(lr){11-12} \cmidrule(lr){14-15}
 &  & PC-6 & 66.72 & \multicolumn{1}{c|}{} & \multicolumn{1}{c|}{} &  & \multicolumn{1}{c|}{PC-6} & \multicolumn{1}{c|}{78.23} & \multicolumn{1}{c|}{} & \multicolumn{1}{c|}{PC-6} & \multicolumn{1}{c|}{55.57} & \multicolumn{1}{c|}{} & \multicolumn{1}{c|}{PC-6} & \multicolumn{1}{c|}{72.24} &  \\ \cmidrule(lr){3-4} \cmidrule(lr){8-9} \cmidrule(lr){11-12} \cmidrule(lr){14-15}
 &  & PC-7 & 71.20 & \multicolumn{1}{c|}{} & \multicolumn{1}{c|}{} &  & \multicolumn{1}{c|}{PC-7} & \multicolumn{1}{c|}{82.71} & \multicolumn{1}{c|}{} & \multicolumn{1}{c|}{PC-7} & \multicolumn{1}{c|}{59.32} & \multicolumn{1}{c|}{} & \multicolumn{1}{c|}{PC-7} & \multicolumn{1}{c|}{77.51} &  \\ \cmidrule(lr){3-4} \cmidrule(lr){8-9} \cmidrule(lr){11-12} \cmidrule(lr){14-15}
 &  & PC-8 & 74.94 & \multicolumn{1}{c|}{} & \multicolumn{1}{c|}{} &  & \multicolumn{1}{c|}{PC-8} & \multicolumn{1}{c|}{86.37} & \multicolumn{1}{c|}{} & \multicolumn{1}{c|}{PC-8} & \multicolumn{1}{c|}{62.55} & \multicolumn{1}{c|}{} & \multicolumn{1}{c|}{PC-8} & \multicolumn{1}{c|}{81.91} &  \\ \cmidrule(lr){3-4} \cmidrule(lr){8-9} \cmidrule(lr){11-12} \cmidrule(lr){14-15}
 &  & PC-9 & 78.07 & \multicolumn{1}{c|}{} & \multicolumn{1}{c|}{} &  & \multicolumn{1}{c|}{PC-9} & \multicolumn{1}{c|}{89.25} & \multicolumn{1}{c|}{} & \multicolumn{1}{c|}{PC-9} & \multicolumn{1}{c|}{65.28} & \multicolumn{1}{c|}{} & \multicolumn{1}{c|}{PC-9} & \multicolumn{1}{c|}{85.74} &  \\ \cmidrule(lr){3-4} \cmidrule(lr){8-9} \cmidrule(lr){11-12} \cmidrule(lr){14-15}
\multirow{-10}{*}{40} & \multirow{-10}{*}{25} & PC-10 & 81.14 & \multicolumn{1}{c|}{\multirow{-10}{*}{3}} & \multicolumn{1}{c|}{\multirow{-10}{*}{2}} & \multirow{-10}{*}{10} & \multicolumn{1}{c|}{PC-10} & \multicolumn{1}{c|}{91.70} & \multicolumn{1}{c|}{\multirow{-10}{*}{7}} & \multicolumn{1}{c|}{PC-10} & \multicolumn{1}{c|}{67.74} & \multicolumn{1}{c|}{\multirow{-10}{*}{19}} & \multicolumn{1}{c|}{PC-10} & \multicolumn{1}{c|}{88.76} & \multirow{-10}{*}{8} \\ \bottomrule
\end{tabular}%
}
\end{table}

Moreover, this scenario reflects a real-world application in a high-dimensional setting where \(n < p\). As observed earlier, Maximum Likelihood Estimation tends to select too few PCs, while the Ledoit Wolf method often retains too many. Both Maximum Likelihood Estimation and Ledoit Wolf fail to accurately represent the total variance as expected by the population distribution. On the other hand, the Standardized Pairwise Differences Covariance method, with its low estimation error, effectively selects a number of PCs that closely align with the population's recommendation.

\section{Conclusion}

These contradictions highlight a crucial issue in PCA: the number of retained PCs depends heavily on the selection method and sample size, which can lead to misleading conclusions, especially in high-dimensional settings such as healthcare research, genomics, and finance. The Kaiser-Guttman Criterion tends to favor retaining a larger number of components, whereas the Scree Plot method suggests keeping fewer PCs. Both approaches, however, can be problematic when selecting either too many or too few components. Retaining too many PCs can lead to overfitting, while retaining too few can cause underfitting, potentially undermining the effectiveness of dimensionality reduction. Misinterpretations in healthcare research, in particular, can have serious consequences, as misleading statistical conclusions may affect critical decision-making processes. Therefore, the percent of cumulative variance method provides a more reasonable number of PCs to retain compared to the Kaiser-Guttman and Scree Plot methods. Moreover, in situations where \textit{n} $<$ \textit{p}, the Standardized Pairwise Difference Covariance (SPDC) estimation method offers a better solution for selecting the optimal number of PCs, outperforming other estimation methods by minimizing the overdispersion error in PCA.

\bibliographystyle{plain}
\bibliography{References.bib}

\end{document}